\documentclass[twocolumn,superscriptaddress,preprintnumbers,amsmath,amssymb,prl]{revtex4}

\usepackage{graphicx}
\usepackage{dcolumn}
\usepackage{bm}
\usepackage{indentfirst} 
\usepackage{color}
\usepackage{CJK}
\usepackage{booktabs}
\usepackage{float}
\usepackage[colorlinks,linkcolor=blue,urlcolor=blue,citecolor=blue]{hyperref}
\usepackage{amsthm}

\usepackage{array} 
\usepackage{caption}
\usepackage{tikz,xcolor}
\usetikzlibrary{decorations.pathmorphing}

\usepackage{tikz,xcolor}

\definecolor{lime}{HTML}{A6CE39}
\DeclareRobustCommand{\orcidicon}{
	\begin{tikzpicture}
	\draw[lime, fill=lime] (0,0) 
	circle [radius=0.16] 
	node[white] {{\fontfamily{qag}\selectfont \tiny ID}};
	\draw[white, fill=white] (-0.0625,0.095) 
	circle [radius=0.007];
	\end{tikzpicture}
	\hspace{-2mm}
}
\foreach \x in {A, ..., Z}{%
	\expandafter\xdef\csname orcid\x\endcsname{\noexpand\href{https://orcid.org/\csname orcidauthor\x\endcsname}{\noexpand\orcidicon}}
}
\foreach \x in {A, ..., Z}{%
	\expandafter\xdef\csname orcid\x\endcsname{\noexpand\href{https://orcid.org/\csname orcidauthor\x\endcsname}{\noexpand\orcidicon}}
}

\begin{document}
\begin{CJK*}{UTF8}{gbsn}

\title{Spacetime Discreteness via Consistent Microscopic Measurement}

\author{Weihu Ma(马维虎)\orcidA{}}
    \email[Contact author: ]{maweihu@fudan.edu.cn}
    \affiliation{Key Laboratory of Nuclear Physics and Ion-beam Application (MOE), Institute of Modern Physics, Fudan University, Shanghai 200433, China}
    \affiliation{Shanghai Research Center for Theoretical Nuclear Physics, NSFC and Fudan University, Shanghai 200438, China}
    
\author{Yu-Gang Ma(马余刚)\orcidB{}}
    \email[Contact author: ]{mayugang@fudan.edu.cn}
    \affiliation{Key Laboratory of Nuclear Physics and Ion-beam Application (MOE), Institute of Modern Physics, Fudan University, Shanghai 200433, China}
    \affiliation{Shanghai Research Center for Theoretical Nuclear Physics, NSFC and Fudan University, Shanghai 200438, China}
    \affiliation{School of Physics, East China Normal University, Shanghai 200241, China}

\date{\today} 

\begin{abstract}
The physical origin of spacetime discreteness remains a central open problem in quantum gravity, with most existing approaches relying on specific microscopic structures or model-dependent assumptions. In this letter, spacetime discreteness can arise instead as a consequence of consistent microscopic measurement. By treating infinitesimal spacetime intervals as scale-dependent measurement outcomes rather than predefined geometric entities, we formulate a Micro-Measurement Principle in which spacetime quantum fluctuations are encoded directly in the scaling structure. An equivalent dual representation of microscopic lengths leads to discrete, equidistant measurement outcomes, with the corresponding scaling-deformed uncertainty relation thereby reducing to the standard Heisenberg form. The microscopic lengths are further governed by a geometric renormalization-group flow admitting finite-length fixed points. This construction preserves Lorentz invariance and general covariance without ad hoc cutoffs or symmetry breaking. Our results show that the classical continuous-spacetime description corresponds to an unstable limiting regime, whereas a finite microscopic length and a discrete spacetime structure arise naturally from the fundamental requirements of micro-measurement consistency.
\end{abstract}


\maketitle
\section{Introduction}
Understanding the microscopic structure of spacetime remains a central challenge in quantum gravity. Since early considerations of a fundamental minimal length, it has been widely suggested that the classical spacetime continuum may break down at sufficiently small scales \cite{Garay1995,Bosso2023,Casadio2023,Bosso2022}. While classical general relativity and quantum field theory on curved backgrounds both rely on the assumption of a smooth continuum geometry, quantum fluctuations of spacetime itself render the operational meaning of infinitesimal distance measurement nontrivial. At the Planck scale, the tension between the deterministic geometry of general relativity and the probabilistic nature of quantum mechanics becomes unavoidable \cite{DeWitt1967,Isham1991,Mead1964}, casting doubt on the self-consistency of a continuous spacetime description under microscopic measurement. Existing approaches to spacetime discreteness typically introduce specific microscopic structures, quantization prescriptions, or model-dependent ultraviolet completions, or interpret discreteness as an effective, scale-dependent phenomenon \cite{Rovelli2010,AshtekarLewandowski2004,Ambjorn2005,Bombelli1987,Reuter1998,Polchinski2011,Liberati2015}. However, a satisfactory theory should derive spacetime microstructure from more primitive, operationally well-defined principles. In this Letter, we show that classical continuity corresponds only to an unstable limiting regime, while the fundamental requirements of microscopic measurement and scale consistency lead to a finite microscopic length and a discrete spacetime structure.

We formulate a Micro-Measurement Principle that treats the measurement of infinitesimal spacetime intervals as a dynamical, scale-dependent process. The framework introduces a dynamical definition of microscopic length, a dual description of its fluctuations, and a geometric renormalization-group (RG) flow governing measurement scales within a covariant setting. Spacetime discreteness then arises as a universal outcome of a scale-dynamics rather than a postulate. This construction provides an operational and covariant origin of a finite microscopic length directly rooted in the consistency of microscopic measurement.

This Letter offers a distinct conceptual presentation and physical interpretation of the Micro-Measurement Principle. The core logical structure and concrete example are presented in a self-contained manner in the main text and Appendix A, while supplementary background and expanded reading can be found in Refs. \cite{weihu1,weihu2}.

Throughout this main content, unless explicitly indicated, repeated indices do not imply Einstein summation.

\section{Micro-measurement}
At quantum-gravity scales, the operational meaning of distance must be reconsidered. Measuring an infinitesimal spacetime separation $dx^{\alpha}$ necessarily requires comparison with a fixed reference length $dx_{0}^{\alpha}$, such as the Planck length. In a classical spacetime geometry, the ratio $dx^{\alpha}/dx_{0}^{\alpha}$ would be taken to be infinitesimal, where index $\alpha$ denotes orthonormal frame coordinates. However, quantum fluctuations of spacetime, rooted in the uncertainty principle, render the local geometry intrinsically dynamical. To account for this, we promote the measurement outcome to a scale-dependent quantity and introduce a scaling function $L^{\alpha}(X^{\alpha})$, defined by
\begin{equation}
\frac{dx^{\alpha}}{dx_{0}^{\alpha}} = L^{\alpha} \left( X^{\alpha} \right)
\label{eq:L_def}
\end{equation}
where $X^{\alpha}$ is a dimensionless scale coordinate. Equation~\eqref{eq:L_def} provides an operational definition of microscopic length, quantifying how a fluctuating spacetime interval compares to a fixed reference scale at a given dynamical scale.

Depending on its scale dependence, the scaling function $L^{\alpha}(X^{\alpha})$ admits qualitatively distinct measurement regimes. A scale-independent form corresponds to fluctuation-free measurements; the limiting case $L^{\alpha} \to 0$ recovers the classical continuous spacetime description, while a nonzero constant represents a static but finite microscopic interval. Allowing a linear or more general nonlinear dependence on $X^{\alpha}$ captures genuine scale-dependent fluctuations of spacetime geometry. In particular, nontrivial scaling behavior encodes microscopic structure beyond the classical continuum and provides a consistent framework for describing discrete spacetime fluctuations.

To describe how microscopic length measurements respond to a refinement of scale, we consider the differential
\begin{equation}
dL^{\alpha} = \bar{L}^{\alpha}-L^{\alpha},
\end{equation}
induced by a rescaling transformation $\bar{X}^{\alpha} = \zeta^{\alpha}(X^{\alpha})X^{\alpha}$, such that $\bar{L}^{\alpha} = L^{\alpha}(\bar{X}^{\alpha})$. The corresponding response of the scaling function defines a first-order fluctuation factor
\begin{equation}
a_{\alpha}(X^{\alpha}) = \frac{\bar{L}^{\alpha}/L^{\alpha}-1}{dL^{\alpha}/dX^{\alpha}},
\label{eq:a_def}
\end{equation}
which satisfies
\begin{equation}
dX^{\alpha} = a_{\alpha}L^{\alpha}.
\label{eq:dX_aL}
\end{equation}
The quantity $a_{\alpha}$ is a dimensionless, direction-dependent amplitude characterizing the local response of microscopic length measurements to changes in scale. It transforms as a Lorentz scalar and does not participate in index contraction unless explicitly promoted to an internal coordinate. The explanation is given in Appendix A.

The scale-dependent Lorentz-invariant line element then be written as
\begin{equation}
\tau^2 = g_{\alpha\beta}L^{\alpha}L^{\beta},
\end{equation}
with the Einstein summation convention implied.
The physical spacetime metric $g_{\alpha\beta}$ is obtained by an anisotropic rescaling of a smooth substrate metric $\hat{g}_{\alpha\beta}$,
\begin{equation}
g_{\alpha\beta} = \hat{g}_{\alpha\beta}\, \frac{a_{\alpha}a_{\beta}}{a^{2}},
\label{eq:g_from_hatg}
\end{equation}
where $a$ (without index) denotes the scalar magnitude inferred from microscopic proper-length measurement $\tau=d\lambda/d\lambda_0$. In this construction, geometric fluctuations enter the physical metric exclusively through the local scaling amplitudes $a_{\alpha}$, while $\hat{g}_{\alpha\beta}$ defines the substrate scale-coordinate manifold.

A second level of fluctuation is characterized by the quantity
\begin{equation}
b_{\alpha} =-\frac{\mathrm{d} a_{\alpha}}{\mathrm{d} X^{\alpha}},
\end{equation}
which measures the nonuniformity of the first-order scale deformation. This induces a higher-level metric $\tilde{g}_{\alpha\beta}$ on the amplitude space, related to the substrate metric through
\begin{equation}
\hat{g}_{\alpha\beta} = \tilde{g}_{\alpha\beta} \frac{b_{\alpha} b_{\beta}}{b^{2}}.
\end{equation}
The resulting hierarchical structure may be summarized as
\begin{equation}
\tilde{g}_{\alpha\beta} \xrightarrow{b_{\alpha}} \hat{g}_{\alpha\beta} \xrightarrow{a_{\alpha}} g_{\alpha\beta},
\label{eq:hierarchy}
\end{equation}
showing that the physical spacetime metric arises from two successive anisotropic scale transformations.

Differential operators in the fluctuating geometry are expressed in terms of the scale coordinates $X^{\alpha}$. In particular,
\begin{equation}
\frac{\partial}{\partial x^{\nu}}
=\frac{dX^{\nu}}{dx^{\nu}}\frac{\partial}{\partial X^{\nu}}
=\frac{a_{\nu}}{dx_{0}^{\nu}}\frac{\partial}{\partial X^{\nu}},
\label{eq:firstpartial}
\end{equation}
and
\begin{equation}
\frac{\partial^{2}}{\partial x^{\mu}\partial x^{\nu}}
=\frac{1}{dx_{0}^{\mu}dx_{0}^{\nu}}
\left(
a_{\mu}a_{\nu}\frac{\partial^{2}}{\partial X^{\mu}\partial X^{\nu}}
- a_{\mu} b_{\nu}\frac{\partial X^{\nu}}{\partial X^{\mu}}\frac{\partial}{\partial X^{\nu}}
\right),
\end{equation}
which makes explicit how microscopic spacetime fluctuations enter local probes through a scale-dependent differential structure. These relations demonstrate that microscopic spacetime fluctuations are not imposed as external corrections, but are intrinsically embedded in the definition of local differential operators. Consequently, within the present framework, the influence of micro-geometry on local physics is encoded kinematically through the scale-dependent differential structure, rather than introduced as a model-dependent modification.

Therefore, the micro-measurement framework replaces the classical infinitesimal $dx^{\alpha}$ with a dynamical scaling function $L^{\alpha}(X^{\alpha})$. The amplitudes $a_{\alpha}$ encode geometric fluctuations, while the hierarchical metric construction (Eq. \ref{eq:hierarchy}) preserves general covariance at each level of the scale structure. The detailed derivation is provided in Appendix A.

The local scaling function \( L^\alpha(X^\alpha) \) may resemble Weyl’s local scale factor \cite{Weyl1918, Weyl1952}, but the two constructions are conceptually distinct. In Weyl’s 1918 geometry, local rescalings are associated with an independent gauge connection and path-dependent length transport. By contrast, \( L^\alpha(X^\alpha) \) in the present framework is an operationally defined measurement function determined by the Micro-Measurement Principle. It is direction-dependent and anisotropic, and it does not introduce an independent Weyl connection or gauge field. Thus, the present construction is Weyl-like only in the limited sense of local scaling, rather than a Weyl gauge theory.

\section{Scaling-deformed commutator}
Several approaches to quantum gravity, including noncommutative geometry \cite{snyder1947}, generalized uncertainty principles (GUP) \cite{kempf1995,Amati1987,Maggiore1993,Scardigli1999}, and doubly special relativity \cite{magueijo2002}, introduce deformations of the canonical commutation relations near the Planck scale. Here we show that the deformations arise within the micro-measurement framework, without invoking additional assumptions beyond scale-dependent measurement. The discreteness implied by microscopic measurement naturally manifests itself at the level of quantum mechanics, providing a direct link between scale-dependent spacetime structure and deformed canonical commutation relations.

We introduce scale-aware position and momentum operators that incorporate local scaling degrees of freedom. The position operator reads
\begin{equation}
\hat{x}^\alpha = \hat{L}^\alpha(X^\alpha) dx_0^\alpha + \hat{x}_0^\alpha,
\label{position}
\end{equation}
and the momentum operator is expressed through the scale-aware derivative
\begin{equation}
\hat{p}_\alpha = -i\hbar \frac{\partial}{\partial x^\alpha}
               = -i r_\hbar \, a_\alpha \, p_\alpha^0 \frac{\partial}{\partial X^\alpha}.
\label{momentum}
\end{equation}
Here \(p_\alpha^0\) is defined as reference momentum and \(r_\hbar = \hbar/(dx_0^\alpha p_\alpha^0)\) a dimensionless normalization.
Evaluating the commutator \([\hat{x}^\alpha,\hat{p}_\alpha]\) yields
\begin{equation}
[\hat{x}^\alpha,\hat{p}_\alpha] = i\hbar \, a_\alpha \frac{d\hat{L}^\alpha}{dX^\alpha}.
\end{equation}
This expression can be recast in the compact form by a scaling-deformed factor (SDF)
\begin{equation}
\hat\vartheta(X^\alpha) = a_\alpha \frac{d\hat{L}^\alpha}{dX^\alpha}=\left(\frac{\hat{\bar{L}}^\alpha}{\hat{L}^\alpha}-1\right); \quad[\hat{x}^\alpha,\hat{p}_\alpha] = i\hat\vartheta\hbar.
\end{equation}
Thus, the usual constant \(\hbar\) is scaled by SDF \(\hat\vartheta=\hat\vartheta(X^\alpha)\) that encodes the local fluctuation structure and induces a family of Hopf-algebra deformations of the Heisenberg algebra \cite{Arzano2008}.
The associated uncertainty relation takes the form
\begin{equation}
\Delta x \, \Delta p \ge \frac{\hbar}{2}\bigl|\langle \hat\vartheta\rangle\bigr|,
\end{equation}
whose lower bound is now dynamical, reflecting the underlying geometric fluctuations. 

In the classical quantum regime,  \(\hat{\bar L}^{\alpha} = 2\hat{L}^{\alpha}\),
one recovers \(\hat\vartheta \to 1\), while the underlying geometry retains a discrete imprint in the form of equidistant spacetime quantization, characterized by increases or decreases in length fluctuations in discrete, equal steps. By contrast, in the static measurement regime with \(\hat{\bar L}^{\alpha} = L^{\alpha} = \text{const}\), the scaling-deformed factor vanishes, \(\hat\vartheta \to 0\), and the position-momentum commutator becomes trivial, reducing the algebra to a commutative one corresponding to a static geometry, which recovers scale invariance. The classical continuous spacetime limit is recovered only in the further limit \(\hat{\bar L}^{\alpha} = L^{\alpha} \to 0\). 

Besides these two limiting cases, scaling-deformed factor \(\hat\vartheta\) acquires a genuinely dynamical meaning. It no longer represents a fixed quantum of action, nor does it vanish identically, but instead parametrizes a scale-dependent deformation of the canonical phase-space algebra induced by microscopic geometric fluctuations. In this regime, nontrivial scale evolution of the microscopic length \(L^\alpha(X^\alpha)\) leads to a continuous interpolation between commutative and canonical quantum structures. The resulting deformed commutation relations encode the local response of spacetime geometry to scale refinement, reflecting an intrinsic coupling between quantum uncertainty and the measurement-induced microstructure. From this perspective, \(\hat\vartheta\) is a geometrically governed quantum variable. It encodes the strength of the scale-deformed Heisenberg algebra arising from scale-dependent spacetime fluctuations, rather than being a universal constant. The deformation of the commutator is therefore a direct manifestation of microscopic measurement consistency in a fluctuating geometry, providing a covariant and non-phenomenological realization of minimal-length physics.

{Applying the scale-aware position and momentum operators from Eqs. (\ref{position}) and (\ref{momentum}) to the standard harmonic oscillator Hamiltonian, we obtain a one-dimensional scaled Schr\"odinger equation that naturally takes the form of a position-dependent-mass (PDM) equation \cite{PDM1,PDM2}. The equation reads
\begin{equation}
-\frac{r_{\hbar}^2}{2r_m}\Bigl[a^{2}\psi''+aa'\psi'\Bigr]+\frac{1}{2}r_m r_\omega^{2}L^{2}\psi = r_E\psi,
\label{eq:PDM}
\end{equation}
where $r_m=m/m_0$, $r_\omega=\omega/\omega_0$, and $r_E=E/E_0$, with $m_0,\;\omega_0$, and $E_0$ are references of the mass, frequency, and energy, respectively. Through wave function renormalization $\psi = a^{-1/2} \phi$, Eq. (\ref{eq:PDM}) is transformed into an equation containing only second-order derivatives.
\begin{equation}
\begin{aligned}
&\frac{-\;r_{\hbar}^2}{2 m_{\text{scaled}}} \frac{d^{2} \phi}{d X^{2}}+\left(\frac{1}{2}m_{\text{scaled}} \omega_{\text{scaled}}^{2} L^{2}+V_{\text{scaled}}\right) \phi=r_E \phi,\\
&\qquad  \text{with  } V_{\text{scaled}}=\frac{r_{\hbar}^2}{4 m_{\text{scaled}}}\left[\frac{ a^{\prime \prime}}{a}-\frac{1}{2} \left(\frac{a^{\prime }}{a}\right)^2\right], 
\label{eq:renormalization}
\end{aligned}
\end{equation}
where $m_{\text{scaled}}=r_m / a^{2}$ and $\omega_{\text{scaled}}=ar_\omega$ are defined as a scale-dependent mass and frequency, respectively. $V_{\text{scaled}}$ is a scale fluctuation-induced potential, providing a non-trivial background. 
It describes the quantum mechanical behavior of a particle moving in a dynamic geometric background, driven by microscopic measurement consistency, in the scale-coordinate $X$ representation. Where the fluctuation amplitude induces a scale-dependent mass \( m_{\text{scaled}}(X)\) and a scale-dependent potential (governed by \( L^2(X)\text{ and }V_{\text{scaled}}(a) \)), governed by scale dynamics. Non-trivial scale fluctuations can generate scale-dependent non-linear geometric effects, even when the commutator recovers canonically Heisenberg ($\vartheta = 1$). This decoupling indicates that nontrivial dynamics can arise from measurement consistency, even when the commutator remains undeformed, distinguishing the present framework from standard GUP. }

{Only when \(L(X)=X\) (linear measurement) and \(a(X)=A+\Lambda X^{2}\) (slow fluctuation), where \(A>0\) is a constant and \(\Lambda\ll1\) a small dimensionless parameter, the scale-deformed commutator of our framework reduces to a GUP-like quadratic correction in the pure coordinate representation.
Expanding \(a(X)\) to first order in \(\Lambda\) gives the energy spectrum
\begin{equation}
E^{dim}_n \approx A\hbar\omega \left(n+\frac12\right) +\frac{A\hbar^2\lambda}{4m}\bigl(2n^{2}+2n+1\bigr) + O(\Lambda^{2}),
\end{equation}
with restored dimensions and \(\lambda = \Lambda/dx_0^2\). $\lambda$ is a characteristic parameter that quantifies the strength of microscopic scale fluctuations, has the dimension of inverse square of length, and directly determines the magnitude of the energy correction $\Delta E \propto \lambda/m$. It implicitly defines a characteristic length scale \( \ell_{\Lambda} \sim 1 / \sqrt{\lambda} \), which sets the scale at which microscopic fluctuations become significant.
The quadratic \(n\)-dependence mirrors that of the KMM GUP \cite{kempf1995}. The opposite mass dependence ($ \propto 1/m $ vs $ \propto m $ in GUP) arises because microscopic fluctuation here has the form in a pure scale coordinate representation, while KMM-GUP introduces a \( p^4 \) term from the deformed commutator when a pure coordinate representation is set. The constant \(A\) merely renormalises the overall energy scale.}

\section{Dual measurement}
A microscopic displacement can be represented either through the nonlinear scaling function or through the rescaled scale increment,  
\begin{equation}  
\mathrm{d}x^{\alpha} = L^{\alpha}(X^{\alpha}) \, \mathrm{d}x_{0}^{\alpha}  
= \mathrm{d}\bar{X}^{\alpha} \, \mathrm{d}x_{0}^{\alpha}.  
\end{equation}  
The nonlinear kernel $L^{\alpha}(X^{\alpha})$ and the linearized differential $\mathrm{d}\bar{X}^{\alpha}$ therefore encode identical fluctuation content, providing two equivalent descriptions of microscopic length measurement. This establishes a dual representation. The scaling function $L^{\alpha}$ captures fluctuations in a scale general-coordinate form, while $\mathrm{d}\bar{X}^{\alpha}$ serves as a locally linearized, fluctuation-adapted variable.  

Under the directional rescaling of the scale coordinate $\bar{X}^\alpha$, we introduce  
\begin{equation}  
\xi^{\alpha} \equiv \frac{\mathrm{d}\bar{X}^{\alpha}}{\mathrm{d}X^{\alpha}}  
= \zeta^{\alpha} + X^{\alpha} \frac{\mathrm{d}\zeta^{\alpha}}{\mathrm{d}X^{\alpha}}. 
\end{equation}  
The mapping between the two representations is then governed by the scale response
\begin{equation}
a_\alpha = \frac{1}{\xi^{\alpha}}, \qquad 
\frac{\mathrm{d}L^{\alpha}}{\mathrm{d}X^{\alpha}}
= \xi^{\alpha}\!\left( \frac{\bar{L}^{\alpha}}{L^{\alpha}} - 1 \right),
\label{eq:mapping}
\end{equation}
so that the first-order fluctuation amplitude satisfies
\( a_\alpha = 1/\xi^\alpha \). This leads to the derivation of
\begin{equation}
\begin{aligned}
&\frac{\partial}{\partial \bar{X}^{\alpha}}=\frac{dX^{\alpha}}{d\bar{X}^{\alpha}}\frac{\partial}{\partial X^{\alpha}}=\frac{1}{\xi^\alpha}\frac{\partial}{\partial X^{\alpha}}=a_\alpha\frac{\partial}{\partial X^{\alpha}},
\end{aligned}
\end{equation}
which is consistent with Eq. (\ref{eq:firstpartial}).

In the coordinate-equivalent representation of the dual measurement, the measured length is related to the coordinate scale through
\begin{equation}
\bar{X}^{\alpha} = \bar X_0^{\alpha} + L^{\alpha}.
\label{eq:linearrepresentation}
\end{equation}
For a reference scale vector\( \bar X_0^{\alpha} \) of finite length $|\bar X_0^{\alpha}|$ and definite direction, inserting Eq. (\ref{eq:linearrepresentation}) into Eq. (\ref{eq:mapping}) gives the discrete doubling condition
\(
\bar{L}^{\alpha}=2L^{\alpha}
\),
and thus \(a_\alpha=\frac{1}{dL^\alpha/dX^\alpha}\).
reflecting the length fluctuations occur in discrete, equidistant steps,
\begin{equation}
\mathrm{d}L^{\alpha} = \bar{L}^{\alpha} - L^{\alpha} = L^{\alpha}.
\end{equation}
At the quantum-mechanical level, this measurement-induced discreteness translates into the above deformed canonical commutation relations, with scaling-deformed factor controlled by the local scale response of $\bar{L}^{\alpha}=2L^{\alpha}$, bridging the description to the classical quantum regime.

Using the rescaling definition 
\(
\bar X^{\alpha}=\zeta^{\alpha}(X^{\alpha})X^{\alpha}
\),
the transformed length reads  
\begin{equation}
\bar L^{\alpha}
=\zeta^{\alpha}(\zeta^{\alpha}X^{\alpha})\,\zeta^{\alpha}X^{\alpha}
-\bar X_0^{\alpha}
=\bar\zeta^{\alpha}\bar X^{\alpha}-\bar X_0^{\alpha},
\end{equation}
where 
\(
\bar\zeta^{\alpha}\equiv\zeta^{\alpha}(\zeta^{\alpha}X^{\alpha})
\).
Imposing the discrete-step condition 
\(
\bar L^{\alpha}=2L^{\alpha}
\)
yields the algebraic constraint equations
\begin{equation}
\bar\zeta^{\alpha}\zeta^{\alpha}
-2\zeta^{\alpha}
+\frac{\bar X_0^{\alpha}}{X^{\alpha}}=0.
\label{eq:neq1_PRL}
\end{equation}

For vanishing reference scale 
\(
\bar X_0^{\alpha}=0
\),
Eq.~(\ref{eq:neq1_PRL}) admits the constant solutions 
\(
\zeta^{\alpha}_*=0
\)
and 
\(
\zeta^{\alpha}_*=2
\).
More generally, any function satisfying the functional constraint 
\(
\zeta^{\alpha}(\zeta^{\alpha}(X^{\alpha}))=2
\)
is also a solution, reflecting a large degeneracy of the zero–reference-scale limit.

A simple nontrivial branch is
\begin{equation}
\zeta^{\alpha}
=2-\frac{\bar X_0^{\alpha}}{X^{\alpha}};\quad \bar\zeta^{\alpha}
=2-\frac{\bar X_0^{\alpha}}{\bar X^{\alpha}}.
\end{equation}
The associated flow exhibits two fixed points:  
(i) a trivial fixed point at 
\(
X^{\alpha}\to\bar X_0^{\alpha}
\)
with 
\(
\zeta^{\alpha}_*=1
\);  
(ii) an infrared fixed point at 
\(
X^{\alpha}\to\infty
\)
with 
\(
\zeta^{\alpha}_*=2
\).
The corresponding microscopic length becomes
\begin{equation}
L^{\alpha}=2X^{\alpha}-2\bar X^{\alpha}_0,
\end{equation}
with constant fluctuation amplitude $a_\alpha=\frac{1}{2}$.

Another branch is
\begin{equation}
\zeta^{\alpha}
=\frac{\bar X_0^{\alpha}}{X^{\alpha}}, 
\qquad 
\bar\zeta^{\alpha}
=\frac{\bar X_0^{\alpha}}{\bar X^{\alpha}},
\end{equation}
which possesses a trivial fixed point 
\(
\zeta^{\alpha}_*=1
\)
at 
\(
X^{\alpha}=\bar X_0^{\alpha}
\).
Here the length collapses identically,
\(
L^{\alpha}\equiv0
\),
and the fluctuation amplitude diverges,
\(
a_\alpha\to\infty
\),
signaling a singular scaling regime.

Beyond these special branches, Eq.~(\ref{eq:neq1_PRL}) admits a rational family of exact solutions.  
With the ansatz
\[
\zeta^{\alpha}(X^\alpha)=\frac{A}{X^\alpha}+B,
\]
consistency requires
\[
A=-B(B-2), 
\qquad 
\bar X_0^{\alpha}=B(B-2)^2.
\]
Thus
\begin{equation}
\zeta^{\alpha}(X^\alpha)
= B-\frac{B(B-2)}{X^\alpha}
\end{equation}
is an exact solution with $B$ satisfying the following cubic relation,
\begin{equation}
B^3-4B^2+4B-\bar X_0^{\alpha}=0,
\end{equation}
organizes the solution space.  
The associated length function becomes
\begin{equation}
L^{\alpha}
=BX^{\alpha}-B(B-2)-\bar X^{\alpha}_0,
\end{equation}
with fluctuation amplitude $a_\alpha=\frac{1}{B}$.
The cubic structure therefore encodes a branching of geometric phases controlled by the reference scale.

More generally, defining the iterative sequence
\[
y_0=X^\alpha,\quad 
y_1=\zeta^\alpha(y_0),\quad 
y_2=\zeta^\alpha(y_1),\dots,
\]
Eq.~(\ref{eq:neq1_PRL}) is equivalent to the second-order recurrence
\[
y_{n+2}
=2-\frac{\bar X_0^{\alpha}}{y_n y_{n+1}}.
\]
Solutions can thus be constructed orbit by orbit: specifying initial data 
\(
(y_0,y_1)
\)
generates a sequence, and defining 
\(
\zeta^\alpha(y_n)=y_{n+1}
\)
yields a function solving the original constraint.  
Fixed points of this iteration sequence correspond to self-consistent scale configurations and can generally be determined by the following self-invariant analysis.

In a regime where successive rescalings are operationally indistinguishable,
which allows one to identify
\begin{equation}
\bar\zeta^{\alpha}=\zeta^{\alpha}(\bar X^{\alpha})=\zeta^{\alpha}(\zeta^{\alpha}(X^{\alpha})X^{\alpha})
=\zeta^{\alpha}(X^{\alpha}),
\end{equation}
which implies that \( \zeta \) is self-invariant along the entire self-rescaling orbit. By iteration, one immediately obtains 
\begin{equation}
\zeta^{\alpha}([\zeta^{\alpha}(X^{\alpha})]^n X^{\alpha}) = \zeta^{\alpha}(X^{\alpha})
\end{equation}
for all \( n \in \mathbb{N}\).
$\zeta^{\alpha}$ is an idempotent element under the induced rescaling composition. It arises when rescalings saturate, and it enforces an algebraically closed scale structure with concrete dynamical and observational consequences. Idempotent elements under induced rescaling compositions commonly appear in physics, where they are associated with RG fixed points, saturated coarse-graining, stable measurement outcomes, and self-similar phases.
This rescaling constraint therefore admits a closed equation,
\begin{equation}
(\zeta^{\alpha})^2-2\zeta^{\alpha}+\frac{\bar X_0^{\alpha}}{X^{\alpha}}
=0,
\label{eq:constraint}
\end{equation}
which characterizes rescaling saturation and admits two solution branches,
\begin{equation}
\zeta_{\pm}^{\alpha}=
1\pm\sqrt{1-\frac{\bar X_0^{\alpha}}{X^{\alpha}}}.
\end{equation}
In this case, $\bar{\zeta}^{\alpha} = \zeta^{\alpha}$, yields two types of fixed points: the trivial one ($X^\alpha\to\bar X^\alpha_0$) with $\zeta^{\alpha}_* = 1$, and the infrared fixed point ($X^{\alpha} \to \infty$) where $ \zeta^{\alpha}_* = 2$ or 0.
Consequently, the corresponding length functions
\begin{equation}
L_{\pm}^{\alpha}=\zeta_{\pm}^{\alpha}X^{\alpha}-\bar X_0^{\alpha}.
\end{equation}
The condition $\bar{\zeta}^{\alpha} = \zeta^{\alpha}$ defines a rescaling-saturated flow.
In the two solutions, when the reference scale vanishes (\(\bar{X}_0^{\alpha} = 0\)), one solution branch reduces to \(\zeta^{\alpha} = 0\), corresponding to the classical continuous spacetime limit with no operationally resolvable microscopic length. The other branch gives \(\zeta^{\alpha} = 2\), yielding \(L^{\alpha} = 2X^{\alpha}\), which approaches the same continuum limit as \(X^{\alpha} \to 0\) and reproduces the ultraviolet-divergent behavior of classical geometry. In contrast, for any nonzero \(\bar{X}_0^{\alpha}\), both branches result in finite microscopic lengths that evolve nontrivially with the scale coordinate, signaling a transition to a fluctuating phase characterized by discrete spacetime intervals.

In this framework, the reference scale \( \bar{X}_0^{\alpha} \) acts as a control parameter, while the scaling function \( L^{\alpha} \) plays the role of an order parameter distinguishing continuous and discrete geometric phases. The transition is intrinsically nonperturbative; any finite \( \bar{X}_0^{\alpha} \) drives the system away from the classical continuum and generates a natural ultraviolet cutoff. Spacetime discreteness thus arises dynamically from the consistent implementation of microscopic scaling measurement, rather than being imposed as an external assumption.

Having identified the regime in which successive rescalings become operationally indistinguishable, we organize the dynamical scale dependence of microscopic lengths via a geometric renormalization-group flow, distinct from conventional dynamical-coupling-strength RG flows \cite{Wetterich1993}. The scaling function $L^{\alpha}(X^{\alpha})$ encodes how microscopic distances vary with the scale $X^{\alpha}$, and its evolution is characterized by the RG $\beta$-function. Further illustrative details and supplementary technical material can be found in Ref. \cite{weihu1},
\begin{equation}
\beta\!\left(L^{\alpha}\right)=X^{\alpha}\frac{dL^{\alpha}}{dX^{\alpha}},
\end{equation}
which measures the response of the microscopic length to logarithmic changes of the scale. Fixed points of the flow, defined by $\beta(L^{\alpha})=0$, correspond to scale-invariant regimes.

Solving the geometric RG flow reveals a nontrivial structure of fixed points governing the microscopic geometry. In the ultraviolet limit $ |X^{\alpha}|\to 0 $, the flow approaches finite fixed points at which the microscopic length remains nonzero, $ L^{\alpha}_{*}=-\bar{X}^{\alpha}_{0} $, while the fluctuation amplitude $ a_{\alpha} $ vanishes algebraically as $ |X^{\alpha}|^{1/2} $. This behavior defines a discrete microscopic phase characterized by suppressed short-distance fluctuations and an intrinsic minimal length. In the opposite infrared limit $ |X^{\alpha}|\to\infty $, the flow converges to fixed points where the scaling function becomes static, $ dL^{\alpha}/dX^{\alpha}=0 $, with $ L^{\alpha}_{*}=-\bar{X}^{\alpha}_{0}/2 $. These infrared fixed points describe a stable discrete spacetime phase in which continuous translational symmetry is effectively reduced to a discrete one.

By contrast, when the reference scale vanishes, $ \bar{X}^{\alpha}_{0}=0 $, the flow admits a trivial fixed point at $ L^{\alpha}_{*}=0 $, corresponding to the classical continuous spacetime limit. This point is unstable; any infinitesimal reference scale drives the system away from continuity toward one of the finite-length discrete phases.
The RG flow therefore organizes microscopic spacetime geometries into distinct universality classes. A central result is that a nonzero reference scale $ \bar{X}^{\alpha}_{0} $ inevitably directs the flow toward a discrete fixed point, in both the ultraviolet and infrared regimes.

\section{Covariance and pre-geometric vacuum}
Any discussion of the microscopic discreteness of spacetime necessarily entails a careful examination of covariance, in particular Lorentz invariance and general covariance.

This framework is constructed to preserve the fundamental symmetries of physics, with its fundamental consistency condition being the universal gauge that all inertial observers employ the same fixed reference length for local measurements. This ensures a consistent operational foundation across frames.

We define a scale-coordinate transformation as a change $X^{\mu}\to X^{\prime\mu}$, which correspondingly induces a spacetime coordinate transformation $x^{\mu}\to x^{\prime\mu}$. Within this setup, the scaling functions $L^{\mu}$ transform as Lorentz vectors, while the fluctuation amplitudes $a_{\mu}$---along with $\zeta^\mu$ and $\xi^\mu$---are Lorentz invariant, manifesting as direction-labeled scalar weights encoding the spectral content of intrinsic quantum fluctuations. The scale coordinates $X^{\mu}$ also transform as vectors, and the composite operator $a_{\mu}\partial/\partial X^{\mu}$ behaves covariantly.
The geometry is organized hierarchically. The physical metric $g_{\mu\nu}$ is constructed from a smooth substrate metric $\hat{g}_{\mu\nu}$ via an anisotropic scaling by the invariant factors $a_{\mu}$. This first-order scaled metric $\hat{g}_{\mu\nu}$ transforms as a tensor under general scale coordinate changes. The second-order factors $b_{\mu}$ lead to a higher-order metric $\tilde{g}_{\mu\nu}$ that remains invariant under the general scale-coordinate transformations. See Appendix A for proof details.

Thus, the micro-measurement principle introduces a refined, scale-dependent geometry without sacrificing Lorentz invariance or general covariance. The fluctuation amplitudes $a_{\mu}$ and their derivatives $b_{\mu}$ encode the spacetime fluctuation spectrum, and the entire hierarchical construction, from scaling functions to scaled metrics, transforms properly under changes of frame. As a result, the spacetime discreteness and minimal-length behavior are observer-independent physical consequences of quantum measurement at microscopic scales within the present framework.

The hierarchical construction reveals that physical metric \(g_{\mu\nu}\) can be obtained from metric \(\tilde{g}_{\mu\nu}\) through successive anisotropic scalings governed by the fluctuation variables \(a_\mu\) and \(b_\mu\). Crucially, while \(g_{\mu\nu}\) depends on the choice of scale and measurement, \(\tilde{g}_{\mu\nu}\) is defined on the manifold of scale fluctuations themselves and is invariant under scale-coordinate transformations. 

We interpret \(\tilde{g}_{\mu\nu}\) as a pre-geometric vacuum. It does not describe distances between events in spacetime; rather, it encodes the geometry of scale fluctuations. In this way, points in the amplitude space labeled by \(a_\mu\) represent configurations of first-order quantum fluctuations, and \(\tilde{g}_{\mu\nu}\) measures the ``distance'' between such configurations. 
Physical spacetime arises as an excitation on top of this vacuum when scale inhomogeneities are switched on. Thus, spacetime and its dynamics are reconstructed phenomena, while \(\tilde{g}_{\mu\nu}\) supplies the minimal, background-free substrate from which all scale-dependent geometries originate. This framework offers a clean separation between the scale-fluctuation content (pre-geometry) and the resulting observable spacetime, providing a novel, covariant route to quantum-gravitational reconstruction.

\section{Conclusion}
In conclusion, the Micro-Measurement Principle provides an operational characterization of microscopic spacetime in terms of a scale-dependent length function \( L^\alpha(X^\alpha) \). The associated scale-deformed uncertainty relation can be reduced to the standard Heisenberg form, indicating that quantum fluctuations are consistently encoded in the scaling structure itself. The existence of an equivalent dual representation implies that microscopic length measurements are intrinsically discrete and equidistant. The resulting geometric renormalization-group flow admits fixed points at which spacetime intervals become finite, scale-invariant, and discrete. Throughout, Lorentz invariance and general covariance are preserved. Spacetime discreteness and minimal length therefore arise as consequences of consistent microscopic measurement, rather than as fundamental assumptions.

\section*{Acknowledgements}
This work is supported by the National Natural Science Foundation of China with Grants No. $12547102$, 12475138 and 12147101, the Strategic Priority Research Program of Chinese Academy of Sciences (No. XDB34000000), and the Science and Technology Commission of Shanghai Municipality (23590780100 and  23JC1400200).

\section{Appendix A: Covariance} 

In this Appendix, unless explicitly indicated, repeated upper-lower index indices imply Einstein summation.

Under a Lorentz transformation \(\Lambda^\mu{}_\nu\), coordinates and derivatives transform in the standard way:
\[
x'^\mu = \Lambda^\mu{}_\nu x^\nu, \quad \partial'_\mu = (\Lambda^{-1})^\nu{}_\mu \partial_\nu.
\]
The framework rests on a single operational postulate: all inertial observers use the \emph{same fixed reference length}:
\[
d\lambda_0 = dx_0^\alpha = dx_0^\beta = dx_0'^\alpha = dx_0'^\beta.
\]
Physical coordinates are defined via scaling functions:
\[
x^\mu = L^\mu(X^\mu)\,dx_0^\mu + x_0^\mu.
\]
Substituting the Lorentz transformation and invoking the invariant reference length yields the transformation law for the scaling functions:
\[
L'^\mu = \Lambda^\mu{}_\nu L^\nu.
\]
Thus \(L^\mu\) transforms as a Lorentz vector.

The scale factors \(a_\mu\) are defined through the scaled derivative
\[
\frac{\partial}{\partial x^\mu} = a_\mu \frac{\partial}{\partial X^\mu} \frac{1}{dx_0^\mu}.
\]
Under Lorentz transformation the left-hand side is covariant. Postulating that the scale coordinates transform as vectors,
\[
X'^\mu = \Lambda^\mu{}_\nu X^\nu,
\]
direct comparison of both sides of the transformed derivative relation immediately gives
\[
a'_\mu = a_\mu.
\]
Hence \(a_\mu\) is a Lorentz scalar. The second-order factors then transform as covectors:
\[
b'_\mu = (\Lambda^{-1})^\nu{}_\mu b_\nu, \qquad b_\mu = -\frac{da_\mu}{dX^\mu}.
\]

To preserve the first-order scaled metric \(\hat{g}_{\alpha\beta} = \eta_{\alpha\beta} \frac{a^2}{a_\alpha a_\beta}\) (Minkowski background), we introduce the scaled Lorentz transformation
\[
\Xi^\alpha{}_\mu = \frac{\partial X^\alpha}{\partial X'^\mu} = \frac{\partial x^\alpha}{\partial x'^\mu} \frac{a_\alpha}{a'_\mu} = \Lambda^\alpha{}_\mu \frac{a_\alpha}{a_\mu}\quad\text{(no sum)}.
\]
Direct substitution shows \(\hat{g}'_{\mu\nu} = \hat{g}_{\mu\nu}\), so \(\Xi\) leaves \(\hat{g}\) Lorentz invariant under scale-coordinate transformation.

General covariance follows from the ordinary tensor transformation law of the physical metric \(g_{\mu\nu}\). For the first-order scaled metric, the chain rule together with \(\frac{\partial x^\alpha}{\partial x'^\mu} = \frac{\partial X^\alpha}{\partial X'^\mu} \frac{a_\mu}{a_\alpha}\) yields
\[
\hat{g}'_{\mu\nu} = \frac{\partial X^\alpha}{\partial X'^\mu} \frac{\partial X^\beta}{\partial X'^\nu} \hat{g}_{\alpha\beta}.
\]
Thus \(\hat{g}_{\mu\nu}\) transforms as a proper tensor under arbitrary scale coordinate changes. For the second-order metric, the amplitude coordinates \(a_\mu\) behave as scalars (\(\partial a_\alpha / \partial a'_\mu = \delta^\alpha_\mu\)), giving
\[
\tilde{g}'_{\mu\nu} = \tilde{g}_{\mu\nu}.
\]
Hence \(\tilde{g}_{\mu\nu}\) is invariant under general scale coordinate transformations.

For further proof details and supplementary material, see Ref. \cite{weihu1}.

\bibliographystyle{unsrt}

\end{CJK*}
\end{document}